\documentclass[
 amsmath,amssymb,
 aps,12pt
]{revtex4-1}

\usepackage{graphicx}
\usepackage{dcolumn}
\usepackage{bm}
\usepackage{color}

\newcommand{\omits}[1]{}

\definecolor{dyellow}{rgb}{1.,0.8,.0}
\definecolor{myblue}{rgb}{.1,.1,.7}
\definecolor{dcyan}{rgb}{.0,.6,.6}
\definecolor{dmagenta}{rgb}{0.6,0.0,0.6}
\definecolor{brown}{rgb}{0.6,0.2,0.}
\definecolor{darkblue}{rgb}{.0,.0,0.5}
\definecolor{darkred}{rgb}{0.75,0.0,0.0}
\definecolor{orange}{rgb}{1.,.6,.0}
\definecolor{dorange}{rgb}{0.8,.4,.0}
\definecolor{green}{rgb}{0.0,1.0,0.0}
\definecolor{darkgreen}{rgb}{0.0,0.7,0.0}
\definecolor{lightgrey}{rgb}{0.7,0.7,0.7}
\definecolor{purple}{rgb}{.4,.0,.4}

\begin{document}

\title{ The entropy of isolated horizons in non-minimally coupling scalar field theory from BF theory}
\author{Jingbo Wang}
\email{wangjb@ihep.ac.cn}
\affiliation{Institute of High Energy Physics and Theoretical Physics Center for
Science Facilities, \\ Chinese Academy of Sciences, Beijing, 100049, People's Republic of China}
\author{Chao-Guang Huang}
\email{huangcg@ihep.ac.cn}
\affiliation{Institute of High Energy Physics and Theoretical Physics Center for
Science Facilities, \\ Chinese Academy of Sciences, Beijing, 100049, People's Republic of China}
 \date{\today}
\begin{abstract}
In this paper, the entropy of isolated horizons in non-minimally coupling scalar field theory
and in the scalar-tensor theory of gravitation is calculated by counting the degree of freedom of quantum states in loop quantum gravity. Instead of boundary Chern-Simons theory, the boundary BF theory is used.  The advantages of the new approaches are
that no spherical symmetry is needed, and that the final result matches exactly with the Wald entropy formula.
\end{abstract}
\maketitle

\section{Introduction}

An isolated horizon (IH) \cite{abf1,afk1} is a generalization of an event horizon of a stationary black hole.  It is a null hypersurface and is determined by its local geometric properties.  As expected, the zero and first law of black hole mechanics for event horizons can be generalized to isolated horizons \cite{abf1,abk1}.  The entropy of an IH also satisfies the Bekenstein-Hawking area law.  The concept of an IH can be generalized to a gravitational theory coupled to matter fields.  For minimally coupled matter fields, an IH is still determined by its local geometric
properties.  Thus, the area-entropy relation remains unchanged when gravity is minimally coupled to matter fields, such as scalar fields, Maxwell fields, Yang-Mills fields \cite{afk1,matter1,matter2}.  For non-minimally coupled matter fields,
the additional condition
is required to define an IH, that the matter field should be time independent on the isolated horizon \cite{nonmin1}.  The entropy of a black hole or
an IH will also depend on the matter fields on the horizon \cite{wald1,wald2}.
For example, if a scalar field $\phi$ is coupled to gravity through the action
\setcounter{equation}{0}
\begin{equation}\label{1}
    S_{\rm NMC}[g_{ab}, \phi]=\int d^4 x \sqrt{-g}[\frac{1}{16\pi G} f(\phi) R-\frac{1}{2} g^{ab}\partial_a \phi \partial_b \phi-V(\phi)],
\end{equation}
where $R$ is the Ricci scalar of the metric $g_{ab}$, $V$ is the potential of the scalar field,
and $f(\phi)$ is an arbitrary function of $\phi$, the entropy of a black hole in the theory is given by the Wald entropy formula
\begin{equation}\label{2}
    S=\frac{1}{4 G } \oint f(\phi) \tilde{\epsilon},
\end{equation}
where the integral is taken on any cross-section of the horizon, and $\tilde{\epsilon}$ is the area 2-form.

On the other hand, the above action is similar to the action of the scalar-tensor theory
of gravitation \cite{{stt1},{stt2},{stt3}},
\begin{equation}\label{2a}
    S_{\rm STT}[g_{ab}, \phi]=\int d^4 x \sqrt{-g}[\frac{1}{16\pi G} \phi R-\frac{\omega(\phi)}{\phi} g^{ab} \partial_a \phi \partial_b \phi-V(\phi)].
\end{equation}
After some transformation, the action (\ref{1}) can be rewritten as the form of (\ref{2a}). Scalar-tensor theory is a popular alternative of general relativity. In particular, some models of scalar-tensor theory can explain the accelerating expansion of the Universe
and the rotation curves of galaxies \cite{{de1},{de2},{dm1},{dm2}}. Recently the scalar-tensor theory has been quantized by the loop quantum method \cite{{zm1},{zm2}}.

The statistical explanation of the entropy for isolated horizons is an important
achievement of loop quantum gravity \cite{{rov},{thie},{al1},{hmh}}.  In Refs. \cite{{abck},{abk1}}, the boundary degrees of freedom are suggested to be described effectively by a Chern-Simons theory. The entropy of the spherical horizon in non-minimally coupling scalar fields has been calculated in Ref. \cite{nonmin2}. The scalar field will modify the level $k$ of the Chern-Simons theory and so the final entropy. Due to the spherical symmetry, Eq. (\ref{2}) reduces to
\begin{equation}\label{2b}
    S=\frac{f(\phi_0) a_0}{4 G },
\end{equation}
which $\phi_0$ is the value of the scalar field on the IH, and $a_0$ is the area of the IH.

Recently, two new closely related but different approaches to interpret the area
entropy statistically in loop quantum gravity have been proposed with the use of boundary BF theory instead of the boundary Chern-Simons theory. With the two approaches, the entropy of an arbitrary isolated horizons in 4 and higher dimensional, pure gravitational theory have been
calculated and the same results are obtained in the two approaches \cite{{wmz},{wh2},{wh1},{hw4}}. In the present paper, the approaches are applied to the IHs in the gravitational theory non-minimally coupling to a scalar field. It will show that the entropy is exactly the Wald entropy formula (\ref{2}). The same procedures may
be applied to the scalar-tensor theory of gravitation and explain the Wald entropy of an IH in the theory.

The paper is organized as follows. In Sec. II, the boundary symplectic form of gravitational theory on an IH as an internal boundary is analyzed.  In Sec. III, the entropy of the isolated horizon is obtained and is shown to match the Wald entropy formula.  Some discussions
are made in Sec. IV.  The paper is in units $c=\hbar=1$.

\section{The symplectic forms}
Parallel to the treatment of pure gravitational theory in 4 dimension \cite{wmz},
consider the first order action \cite{nonmin1} of (\ref{1})
\begin{equation}\label{3}
    S_{\rm NMC}[e, A, \phi]=\int_{\mathcal{M}} -\frac{1}{2\kappa} f(\phi) \Sigma_{IJ}   \wedge F^{IJ}+\frac{1}{2} K(\phi) (*{\rm d} \phi)\wedge {\rm d} \phi-V(\phi) \epsilon,
\end{equation}
where $\kappa\equiv 8\pi G$, ${\cal M}$ is a 4-dimensional asymptotically flat region bounded by an IH, as its internal boundary,
\begin{equation}\label{4}
\Sigma_{IJ}=\frac 1 2\varepsilon_{IJKL} e^K\wedge e^L,
\end{equation}
$e^I$ is the orthogonal co-tetrad, $F^{IJ}={\rm d}A^{IJ}+A^I{}_K\wedge A^{KJ}$
is the curvature of the SO$(3,1)$ connection 1-form $A^{IJ}$, $*$ is the Hodge dual, $\epsilon$ is the volume 4-form on $\mathcal{M}$ and
\begin{equation}\label{5}
    K(\phi)=1+\frac{3}{2 \kappa}\frac{(f'(\phi))^2}{f(\phi)}.
\end{equation}
The symplectic current of the theory is \cite{nonmin2}
\begin{equation}\label{10a}
J(\delta_1,\delta_2)=\frac{1}{\kappa} \delta_{[1} (f(\phi) \Sigma_{IJ})   \wedge \delta_{2]} A^{IJ}+2 K(\phi) \delta_{[1}(*{\rm d} \phi)\wedge \delta_{2]}\phi.
\end{equation}

For isolated horizons, the scalar field should satisfy \cite{nonmin1}
\begin{equation}\label{6}
    \mathcal{L}_l \phi \triangleq 0,
\end{equation}
where $l$ is the normal vector of the IH.  This is the additional condition which means that the scalar field is time independent on the isolated horizon $\Delta$. Hereafter equalities on $\Delta$ will be denoted by the symbol $\triangleq$.

As in Ref. \cite{wmz}, the following set of co-tetrad fields is chosen
\begin{equation}\begin{split}\label{7}
    e^0=\sqrt{\frac{1}{2}}(\alpha n+\frac{1}{\alpha} l),\ &e^1=\sqrt{\frac{1}{2}}(\alpha n-\frac{1}{\alpha} l),\\
    e^2=\sqrt{\frac{1}{2}}(m+\bar{m}),\quad &e^3=i\sqrt{\frac{1}{2}}(m-\bar{m}),
\end{split}\end{equation}
where $\alpha(x)$ is an arbitrary function of the coordinates, and $(l, n, m, \bar{m})$ is the Newman-Penrose null co-tetrad.
Restricted to the horizon $\Delta$, the revelent co-tetrad fields (\ref{7}) satisfies
\begin{equation}\label{8}
    e^0\triangleq e^1.
\end{equation}
After some straightforward calculation, the following condition can be obtained:
\begin{equation}\label{9}\begin{split}
&\Sigma_{0i}\triangleq -\Sigma_{1i},\quad A^{0i}\triangleq A^{1i},\quad \forall i=2,3,\\
&A^{01}\triangleq (\tilde{\kappa} {\rm d}v+{\rm d}(\ln\alpha))+(\pi m+\bar{\pi} \bar{m}):\triangleq \bar{A}^{01}+\tilde{A}^{01},
\end{split}\end{equation}
where $ \tilde{\kappa}$ is the surface gravity, $\pi,\bar{\pi}$ are spin coefficients,
and $\bar A^{01}$ and $\tilde A^{01}$ are the nonrotating and rotating parts of $A^{01}$, respectively.  The first equation of (\ref{9}) gives, straightly,
\begin{equation}\label{10}
f(\phi) \Sigma_{0i}\triangleq -f(\phi) \Sigma_{1i},\quad \forall i=2,3.
\end{equation}

With the help of these relations, the integral of the symplectic current (\ref{10a}) over
the IH
can be reduced to
\begin{equation}\label{11}
\Omega_{\Delta}(\delta_1,\delta_2)=\frac{2}{\kappa}\int_{\Delta} \delta_{[1}(f(\phi) \Sigma_{01})\wedge\delta_{2]} \bar A^{01}.
\end{equation}
This is because
the surface symplectic structure only depends on the gravitational part, and because the contribution of the rotating part $\tilde{A}^{01}$ to the integral of the symplectic current over the IH is zero though it does not vanish for a generic IH.
The proof of the latter fact is similar to the pure Einstein gravity \cite{wh2},
with the modification of the definition of conserved angular momentum \cite{nonmin1},
\begin{equation}\label{11a}
    J[w]=-\frac{1}{\kappa}\int_H (w \lrcorner  \tilde A^{01})f\Sigma_{01},
\end{equation}
where $H$ is the section of the IH, and $w$ is an arbitrary vector on $H$.

Notice that, due to the condition (\ref{6}), it is easy to show that $f(\phi)\Sigma_{01}$
is a closed 2-form on the IH:
\begin{equation}\label{12}
    {\rm d}(f(\phi) \Sigma_{01})=\frac{{\rm d} f(\phi)}{{\rm d} \phi} {\rm d}\phi \wedge \Sigma_{01}+f(\phi){\rm d} \Sigma_{01}\triangleq 0.
\end{equation}
So there always exists a 1-form ${\cal B}$
such that
\begin{equation}\label{13}
    {\rm d} {\cal B} \triangleq \frac{1}{\kappa}f(\phi) \Sigma_{01}
       =:\frac{1}{2\kappa}\Pi^1.
\end{equation}
$\Pi^i=2 f(\phi) \Sigma_{0i}
$ is the conjugate momentum of the connection $A^i$ in the non-minimally coupling theory \cite{nonmin2}.  The integral of (\ref{13}) over a section of the IH is
the flux of $\Pi^1$ over the horizon divided by $\kappa$,
\begin{equation}\label{14}
    \oint_{H} {\rm d}{\cal B}=\frac{1}{\kappa}\oint_{H} f(\phi)\Sigma_{01}:=\frac{\tilde{a}_H}{\kappa}.
\end{equation}

With the use of the ${\cal B}$ field, the Eq.(\ref{11}) can be rewritten as
\begin{eqnarray}\label{14a}
\Omega_{\Delta}(\delta_1,\delta_2)&=&\frac{2}{\kappa}\int_{\Delta} \delta_{[1}(f(\phi) \Sigma_{01})\wedge\delta_{2]} \bar{A}^{01}\nonumber\\
&=&2\int_{H_2} \delta_{[1}{\cal B}\wedge\delta_{2]} \bar{A}^{01}-2\int_{H_1} \delta_{[1}{\cal B}\wedge\delta_{2]} \bar{A}^{01}.
\end{eqnarray}
If the nonrotating part $\bar{A}^{01}$ on the IH is identified to an SO$(1,1)$ connection field ${\cal A}$, (\ref{14a}) takes the form of the difference of the symplectic structure of an SO$(1,1)$
BF theory between two sections of IH.  Hence, the symplectic form for the theory (\ref{3})
reads
\begin{equation}\label{eq:SSfor3}
  \Omega = \int_M \frac{1}{\kappa}\delta_{[1}(f(\phi)\Sigma_{IJ})\wedge\delta_{2]} A^{IJ}+
2 K(\phi) \delta_{[1}(*{\rm d} \phi)\wedge \delta_{2]}\phi + 2\int_{H} \delta_{[1}{\cal B}\wedge\delta_{2]} \bar{A}^{01},
\end{equation}
where $M$ is the 3-dimensional spacelike hypersurface.  It shows that the degrees of freedom in the theory (\ref{3}) are divided into two classes.  One class are the bulk degrees of
freedom and the other are the boundary degrees of freedom.
The boundary degrees of freedom are described by an SO$(1,1)$ BF theory, as in pure gravity theory \cite{{wmz},{wh1},{wh2}}.

As the usual discussion of a theory on a manifold with a boundary, one may also consider the action
\begin{equation}\label{TotAction}
  S_{\rm Tot}=S_{\rm NMC}[e, A, \phi] +\int_{\Delta}{\cal B}\wedge {\rm d} {\cal A}.
\end{equation}
The latter term is the action for an SO$(1,1)$ BF theory on the internal boundary $\Delta$,
in which ${\cal A}$ is the SO$(1,1)$ connection and ${\cal B}$ is the $B$-field in the BF theory.
If ${\cal A}$ is identified with the boundary value of $\bar A^{01}$, the equation of motion for the BF theory on $\Delta$, Eq. (\ref{13}) and ${\rm d}{\cal{A}}\triangleq 0$, can be obtained.
The symplectic form for the theory (\ref{TotAction}) becomes
\begin{equation}\label{eq:SSforTotAction}
  \Omega = \int_M \frac{1}{\kappa} \delta_{[1}(f(\phi)\Sigma_{IJ})\wedge\delta_{2]} A^{IJ}+
2 K(\phi) \delta_{[1}(*{\rm d} \phi)\wedge \delta_{2]}\phi.
\end{equation}
Obviously, only the bulk degrees of freedom remain, just as in \cite{hw4}.

\section{The entropy}
In the previous section, two different symplectic forms for the gravitational theory coupled nonminimally to a scalar field are presented.
In this section, we will calculate the entropy in two different approaches, fitting the above
 two symplectic forms, respectively.
\subsection{The first approach}
The first approach follows the procedure of papers \cite{{wmz},{wh1},{wh2}}. Since the symplectic form (\ref{eq:SSfor3}) contains both
the bulk term and the boundary term, the total Hilbert space after loop quantization is the tensor product of bulk and boundary Hilbert spaces. The boundary Hilbert space for the SO$(1,1)$ BF theory is constructed in \cite{{wmz},{hw4}}. The bulk Hilbert space describes not only the polymer excitations of the geometry but also the excitation of the scalar field \cite{nonmin2}. The $\Pi^i$ and $A^i$ are a pair of canonical variables in Hamiltonian formulation of general relativity.
As the eigenvalues of the gravitational momentum $\Sigma^i$ in vacuum and
in the minimally coupled cases, the eigenvalues of $\Pi^i$, instead of the eigenvalues of the
operator of kinematic area, are discrete.
After loop quantization, the operator of the flux through 2-dimensional bounded neighborhood
$s_p$ associated to the `puncture' $p$ has the following form of the eigenvalues
\begin{equation}\label{16a}\begin{split}
   \oint_{s_p} \hat{\Pi}^1|\{j_p,m_p\};\cdots>=16\pi\gamma l^2_{Pl} m_p|\{j_p,m_p\};\cdots>,
\end{split}\end{equation}
where $j_p,\ m_p$ are the quantum spin quantum number and the magnetic quantum number associated with edges of the spin network, respectively. From this expression, the area spectrum for a spherically symmetric IH coincides with the form of \cite{nonmin2}.  The quantum scalar field takes continuous value at each vertex. Since the degrees of freedom of the scalar field will be traced out, it does not affect the entropy of the IH.

The quantum version of the boundary condition (\ref{13}) is
\begin{equation}\label{16}
    (\textrm{Id}\otimes \oint_{s_p} {\rm d}\hat{\cal B}-\frac{1}{2\kappa}\oint_{s_p} \hat{\Pi}^1\otimes \textrm{Id})(\Psi_v \otimes \Psi_b)=0,
\end{equation}
where Id means the identity operator, and
$\Psi_v$ and $\Psi_b$ bulk and boundary states, respectively.
From the boundary condition (\ref{16}) one can get the relation between the eigenvalues of those two operators
\begin{equation}\label{17}
    a_p=\gamma m_p,\qquad 2m_p\in \mathbb{Z} \setminus \{0\},
\end{equation}
where $a_p$ is the eigenvalue of $\oint_{s_p} {\rm d}\hat{\cal B}$ acting the boundary state, $\gamma$ the Barbero-Immirzi parameter.

\subsection{The second approach}
The second approach follows the paper \cite{hw4}.  The symplectic form (\ref{eq:SSforTotAction}) just contains the bulk term and thus the Hilbert space is just
the bulk Hilbert space whose basis is set up by a series of spin network states with the quantum numbers $(j,\ m)$ on each edge and ($\phi$, intertwiners, $\cdots$) on each vertex.  Even though, the boundary BF theory should still be quantized independently.
Since the boundary BF theory coupled to a bulk gravity can be decomposed classically into a
closed form just like the $B$ field in a pure BF theory and a fixed, non-closed form
determined by the coupling, one may first quantize the SO$(1,1)$ pure BF theory in
loop-quantization formalism \cite{bf1}, setting up the complete basis of the boundary Hilbert space, and then use the quantum version of coupling term [the first equality of (\ref{14})] to choose suitable bulk spin network states via triangulations on a section of the IH defined by boundary `spin-network' states.  For a suitable bulk state, there must be a set of edges starting or ending at 2-simplices of a triangulation on the IH.
The flux over each 2-simplex has the eigenvalue $\kappa \gamma m_p$, where $2m_p \in \mathbb{Z}\
\backslash \ \{0\}$ again.

\subsection{The entropy}
The flux constraint [the second equality of (\ref{14})] can be written as
\begin{equation}\label{18}
  \sum_{p \,\in \,\mathcal{P} \mbox{\,or\,}{\cal S}} |m_p|=a,\quad 2m_p \in \mathbb{Z}\setminus \{0\},
\end{equation}
where ${\cal P}$ and ${\cal S}$ are the set of `punctures' in the first approach and the set of 2-simplices in the second approach, $a=\frac{\tilde{a}_H}{8\pi \gamma l_{Pl}^2}$.
When the topological constraint on $S^2$ is ignored (it will give the sub-sub-leading term), the number of the compatible states is
 \begin{equation}\label{19}
    \mathcal{N}=\sum_{n=1}^{n=2a} C_{2a-1}^{n-1} 2^n=2\times 3^{2a-1},
\end{equation}where $C_i^j$ are the binomial coefficients.
So the entropy is given by
\begin{equation}\label{20}
    S=\ln\mathcal{N}=2a\ln3+\ln \frac{2}{3}=\frac{\ln3}{\pi\gamma}\frac{\tilde{a}_H}{4l^2_{Pl}}+\ln \frac{2}{3}=\frac{1}{4l^2_{Pl}}(\oint_{H} f(\phi)\Sigma_{01})+\ln \frac{2}{3},
\end{equation}
which is just the Wald entropy formula (\ref{2}) plus a constant correction term. The Barbero-Immirzi parameter is chosen to be $\gamma=\ln3/\pi$ as in pure Einstein gravity \cite{wmz}.
\section{Conclusion}
In this paper, the entropy of the isolated horizons in non-minimally coupling scalar field theory is calculated by counting the degree of freedom of quantum states in loop quantum gravity. The entropy matches exactly the Wald entropy formula. In our calculation, it does not need any symmetries for the section of the isolated horizons. The generalization to higher dimension is straightforward \cite{wh1}. The same conclusion is also applicable to
the scalar-tensor theory of gravitation because it has a similar action.

The flux operator which conjugates to the connection is $\Pi^i=2 f(\phi) \Sigma_{0i}$ instead of $\Sigma_{0i}$ in non-minimally coupling scalar field theory.  While $\Sigma_{01}$ gives the area of the horizon, it is the flux $\Pi^1=2 f(\phi) \Sigma_{01}$ that gives the entropy. So the ``flux constraint" (\ref{18}) is a better choice than the ``area constraint" in usual Chern-Simons theory \cite{wh3}.

\section*{Acknowledgments}
This work is supported by National Natural Science Foundation of China under the grant
11275207.
 \bibliography{non1}

\begin{thebibliography}{29}%
\makeatletter
\providecommand \@ifxundefined [1]{%
 \@ifx{#1\undefined}
}%
\providecommand \@ifnum [1]{%
 \ifnum #1\expandafter \@firstoftwo
 \else \expandafter \@secondoftwo
 \fi
}%
\providecommand \@ifx [1]{%
 \ifx #1\expandafter \@firstoftwo
 \else \expandafter \@secondoftwo
 \fi
}%
\providecommand \natexlab [1]{#1}%
\providecommand \enquote  [1]{``#1''}%
\providecommand \bibnamefont  [1]{#1}%
\providecommand \bibfnamefont [1]{#1}%
\providecommand \citenamefont [1]{#1}%
\providecommand \href@noop [0]{\@secondoftwo}%
\providecommand \href [0]{\begingroup \@sanitize@url \@href}%
\providecommand \@href[1]{\@@startlink{#1}\@@href}%
\providecommand \@@href[1]{\endgroup#1\@@endlink}%
\providecommand \@sanitize@url [0]{\catcode `\\12\catcode `\$12\catcode
  `\&12\catcode `\#12\catcode `\^12\catcode `\_12\catcode `\%12\relax}%
\providecommand \@@startlink[1]{}%
\providecommand \@@endlink[0]{}%
\providecommand \url  [0]{\begingroup\@sanitize@url \@url }%
\providecommand \@url [1]{\endgroup\@href {#1}{\urlprefix }}%
\providecommand \urlprefix  [0]{URL }%
\providecommand \Eprint [0]{\href }%
\providecommand \doibase [0]{http://dx.doi.org/}%
\providecommand \selectlanguage [0]{\@gobble}%
\providecommand \bibinfo  [0]{\@secondoftwo}%
\providecommand \bibfield  [0]{\@secondoftwo}%
\providecommand \translation [1]{[#1]}%
\providecommand \BibitemOpen [0]{}%
\providecommand \bibitemStop [0]{}%
\providecommand \bibitemNoStop [0]{.\EOS\space}%
\providecommand \EOS [0]{\spacefactor3000\relax}%
\providecommand \BibitemShut  [1]{\csname bibitem#1\endcsname}%
\let\auto@bib@innerbib\@empty
\bibitem [{\citenamefont {{Ashtekar}}\ \emph {et~al.}(1999)\citenamefont
  {{Ashtekar}}, \citenamefont {{Beetle}},\ and\ \citenamefont
  {{Fairhurst}}}]{abf1}%
  \BibitemOpen
  \bibfield  {author} {\bibinfo {author} {\bibfnamefont {A.}~\bibnamefont
  {{Ashtekar}}}, \bibinfo {author} {\bibfnamefont {C.}~\bibnamefont
  {{Beetle}}}, \ and\ \bibinfo {author} {\bibfnamefont {S.}~\bibnamefont
  {{Fairhurst}}},\ }\href@noop {} {\bibfield  {journal} {\bibinfo  {journal}
  {Class. Quant. Grav.}\ }\textbf {\bibinfo {volume} {16}},\ \bibinfo {pages}
  {L1} (\bibinfo {year} {1999})}\BibitemShut {NoStop}%
\bibitem [{\citenamefont {{Ashtekar}}\ \emph
  {et~al.}(2000{\natexlab{a}})\citenamefont {{Ashtekar}}, \citenamefont
  {{Fairhurst}},\ and\ \citenamefont {{Krishnan}}}]{afk1}%
  \BibitemOpen
  \bibfield  {author} {\bibinfo {author} {\bibfnamefont {A.}~\bibnamefont
  {{Ashtekar}}}, \bibinfo {author} {\bibfnamefont {S.}~\bibnamefont
  {{Fairhurst}}}, \ and\ \bibinfo {author} {\bibfnamefont {B.}~\bibnamefont
  {{Krishnan}}},\ }\href {\doibase 10.1103/PhysRevD.62.104025} {\bibfield
  {journal} {\bibinfo  {journal} {Phys. Rev. D}\ }\textbf {\bibinfo {volume}
  {62}},\ \bibinfo {pages} {104025} (\bibinfo {year} {2000}{\natexlab{a}})},\
  \Eprint {http://arxiv.org/abs/gr-qc/0005083} {arXiv:gr-qc/0005083 [gr-qc]}
  \BibitemShut {NoStop}%
\bibitem [{\citenamefont {{Ashtekar}}\ \emph
  {et~al.}(2000{\natexlab{b}})\citenamefont {{Ashtekar}}, \citenamefont
  {{Baez}},\ and\ \citenamefont {{Krasnov}}}]{abk1}%
  \BibitemOpen
  \bibfield  {author} {\bibinfo {author} {\bibfnamefont {A.}~\bibnamefont
  {{Ashtekar}}}, \bibinfo {author} {\bibfnamefont {J.}~\bibnamefont {{Baez}}},
  \ and\ \bibinfo {author} {\bibfnamefont {K.}~\bibnamefont {{Krasnov}}},\
  }\href@noop {} {\bibfield  {journal} {\bibinfo  {journal} {Adv. Theor. Math.
  Phys.}\ }\textbf {\bibinfo {volume} {4}},\ \bibinfo {pages} {1} (\bibinfo
  {year} {2000}{\natexlab{b}})},\ \Eprint {http://arxiv.org/abs/gr-qc/0005126}
  {arXiv:gr-qc/0005126 [gr-qc]} \BibitemShut {NoStop}%
\bibitem [{\citenamefont {{Corichi}}\ and\ \citenamefont
  {{Sudarsky}}(2000)}]{matter1}%
  \BibitemOpen
  \bibfield  {author} {\bibinfo {author} {\bibfnamefont {A.}~\bibnamefont
  {{Corichi}}}\ and\ \bibinfo {author} {\bibfnamefont {D.}~\bibnamefont
  {{Sudarsky}}},\ }\href {\doibase 10.1103/PhysRevD.61.101501} {\bibfield
  {journal} {\bibinfo  {journal} {Phys. Rev. D}\ }\textbf {\bibinfo {volume}
  {61}},\ \bibinfo {pages} {101501} (\bibinfo {year} {2000})},\ \Eprint
  {http://arxiv.org/abs/gr-qc/9912032} {arXiv:gr-qc/9912032 [gr-qc]}
  \BibitemShut {NoStop}%
\bibitem [{\citenamefont {{Corichi}}\ \emph {et~al.}(2000)\citenamefont
  {{Corichi}}, \citenamefont {{Nucamendi}},\ and\ \citenamefont
  {{Sudarsky}}}]{matter2}%
  \BibitemOpen
  \bibfield  {author} {\bibinfo {author} {\bibfnamefont {A.}~\bibnamefont
  {{Corichi}}}, \bibinfo {author} {\bibfnamefont {U.}~\bibnamefont
  {{Nucamendi}}}, \ and\ \bibinfo {author} {\bibfnamefont {D.}~\bibnamefont
  {{Sudarsky}}},\ }\href {\doibase 10.1103/PhysRevD.62.044046} {\bibfield
  {journal} {\bibinfo  {journal} {Phys. Rev. D}\ }\textbf {\bibinfo {volume}
  {62}},\ \bibinfo {pages} {044046} (\bibinfo {year} {2000})},\ \Eprint
  {http://arxiv.org/abs/gr-qc/0002078} {arXiv:gr-qc/0002078 [gr-qc]}
  \BibitemShut {NoStop}%
\bibitem [{\citenamefont {{Ashtekar}}\ \emph {et~al.}(2003)\citenamefont
  {{Ashtekar}}, \citenamefont {{Corichi}},\ and\ \citenamefont
  {{Sudarsky}}}]{nonmin1}%
  \BibitemOpen
  \bibfield  {author} {\bibinfo {author} {\bibfnamefont {A.}~\bibnamefont
  {{Ashtekar}}}, \bibinfo {author} {\bibfnamefont {A.}~\bibnamefont
  {{Corichi}}}, \ and\ \bibinfo {author} {\bibfnamefont {D.}~\bibnamefont
  {{Sudarsky}}},\ }\href {\doibase 10.1088/0264-9381/20/15/310} {\bibfield
  {journal} {\bibinfo  {journal} {Class. Quant. Grav.}\ }\textbf {\bibinfo
  {volume} {20}},\ \bibinfo {pages} {3413} (\bibinfo {year} {2003})},\ \Eprint
  {http://arxiv.org/abs/gr-qc/0305044} {arXiv:gr-qc/0305044 [gr-qc]}
  \BibitemShut {NoStop}%
\bibitem [{\citenamefont {{Wald}}(1993)}]{wald1}%
  \BibitemOpen
  \bibfield  {author} {\bibinfo {author} {\bibfnamefont {R.~M.}\ \bibnamefont
  {{Wald}}},\ }\href {\doibase 10.1103/PhysRevD.48.R3427} {\bibfield  {journal}
  {\bibinfo  {journal} {Phys. Rev. D}\ }\textbf {\bibinfo {volume} {48}},\
  \bibinfo {pages} {3427} (\bibinfo {year} {1993})},\ \Eprint
  {http://arxiv.org/abs/gr-qc/9307038} {arXiv:gr-qc/9307038 [gr-qc]}
  \BibitemShut {NoStop}%
\bibitem [{\citenamefont {{Jacobson}}\ \emph {et~al.}(1994)\citenamefont
  {{Jacobson}}, \citenamefont {{Kang}},\ and\ \citenamefont {{Myers}}}]{wald2}%
  \BibitemOpen
  \bibfield  {author} {\bibinfo {author} {\bibfnamefont {T.}~\bibnamefont
  {{Jacobson}}}, \bibinfo {author} {\bibfnamefont {G.}~\bibnamefont {{Kang}}},
  \ and\ \bibinfo {author} {\bibfnamefont {R.~C.}\ \bibnamefont {{Myers}}},\
  }\href {\doibase 10.1103/PhysRevD.49.6587} {\bibfield  {journal} {\bibinfo
  {journal} {Phys. Rev. D}\ }\textbf {\bibinfo {volume} {49}},\ \bibinfo
  {pages} {6587} (\bibinfo {year} {1994})},\ \Eprint
  {http://arxiv.org/abs/gr-qc/9312023} {arXiv:gr-qc/9312023 [gr-qc]}
  \BibitemShut {NoStop}%
\bibitem [{\citenamefont {{Brans}}\ and\ \citenamefont {{Dicke}}(1961)}]{stt1}%
  \BibitemOpen
  \bibfield  {author} {\bibinfo {author} {\bibfnamefont {C.}~\bibnamefont
  {{Brans}}}\ and\ \bibinfo {author} {\bibfnamefont {R.}~\bibnamefont
  {{Dicke}}},\ }\href {\doibase 10.1103/PhysRev.124.925} {\bibfield  {journal}
  {\bibinfo  {journal} {Phys. Rev.}\ }\textbf {\bibinfo {volume} {124}},\
  \bibinfo {pages} {925} (\bibinfo {year} {1961})}\BibitemShut {NoStop}%
\bibitem [{\citenamefont {{Bergmann}}(1968)}]{stt2}%
  \BibitemOpen
  \bibfield  {author} {\bibinfo {author} {\bibfnamefont {P.~G.}\ \bibnamefont
  {{Bergmann}}},\ }\href {\doibase 10.1007/BF00668828} {\bibfield  {journal}
  {\bibinfo  {journal} {Int. J. Theor. Phys.}\ }\textbf {\bibinfo {volume}
  {1}},\ \bibinfo {pages} {25} (\bibinfo {year} {1968})}\BibitemShut {NoStop}%
\bibitem [{\citenamefont {{Wagoner}}(1970)}]{stt3}%
  \BibitemOpen
  \bibfield  {author} {\bibinfo {author} {\bibfnamefont {R.~V.}\ \bibnamefont
  {{Wagoner}}},\ }\href {\doibase 10.1103/PhysRevD.1.3209} {\bibfield
  {journal} {\bibinfo  {journal} {Phys. Rev. D}\ }\textbf {\bibinfo {volume}
  {1}},\ \bibinfo {pages} {3209} (\bibinfo {year} {1970})}\BibitemShut
  {NoStop}%
\bibitem [{\citenamefont {{Boisseau}}\ \emph {et~al.}(2000)\citenamefont
  {{Boisseau}}, \citenamefont {{Esposito-Farese}}, \citenamefont {{Polarski}},\
  and\ \citenamefont {{Starobinsky}}}]{de1}%
  \BibitemOpen
  \bibfield  {author} {\bibinfo {author} {\bibfnamefont {B.}~\bibnamefont
  {{Boisseau}}}, \bibinfo {author} {\bibfnamefont {G.}~\bibnamefont
  {{Esposito-Farese}}}, \bibinfo {author} {\bibfnamefont {D.}~\bibnamefont
  {{Polarski}}}, \ and\ \bibinfo {author} {\bibfnamefont {A.~A.}\ \bibnamefont
  {{Starobinsky}}},\ }\href {\doibase 10.1103/PhysRevLett.85.2236} {\bibfield
  {journal} {\bibinfo  {journal} {Phys. Rev. Lett.}\ }\textbf {\bibinfo
  {volume} {85}},\ \bibinfo {pages} {2236} (\bibinfo {year} {2000})},\ \Eprint
  {http://arxiv.org/abs/gr-qc/0001066} {arXiv:gr-qc/0001066 [gr-qc]}
  \BibitemShut {NoStop}%
\bibitem [{\citenamefont {{Boisseau}}(2011)}]{de2}%
  \BibitemOpen
  \bibfield  {author} {\bibinfo {author} {\bibfnamefont {B.}~\bibnamefont
  {{Boisseau}}},\ }\href {\doibase 10.1103/PhysRevD.83.043521} {\bibfield
  {journal} {\bibinfo  {journal} {Phys. Rev. D}\ }\textbf {\bibinfo {volume}
  {83}},\ \bibinfo {pages} {043521} (\bibinfo {year} {2011})},\ \Eprint
  {http://arxiv.org/abs/1011.2915} {arXiv:1011.2915 [astro-ph.CO]} \BibitemShut
  {NoStop}%
\bibitem [{\citenamefont {{Lee}}\ and\ \citenamefont {{Lee}}(2004)}]{dm1}%
  \BibitemOpen
  \bibfield  {author} {\bibinfo {author} {\bibfnamefont {T.~H.}\ \bibnamefont
  {{Lee}}}\ and\ \bibinfo {author} {\bibfnamefont {B.~J.}\ \bibnamefont
  {{Lee}}},\ }\href {\doibase 10.1103/PhysRevD.69.127502} {\bibfield  {journal}
  {\bibinfo  {journal} {Phys. Rev. D}\ }\textbf {\bibinfo {volume} {69}},\
  \bibinfo {pages} {127502} (\bibinfo {year} {2004})}\BibitemShut {NoStop}%
\bibitem [{\citenamefont {{Catena}}\ \emph {et~al.}(2004)\citenamefont
  {{Catena}}, \citenamefont {{Fornengo}}, \citenamefont {{Masiero}},
  \citenamefont {{Pietroni}},\ and\ \citenamefont {{Rosati}}}]{dm2}%
  \BibitemOpen
  \bibfield  {author} {\bibinfo {author} {\bibfnamefont {R.}~\bibnamefont
  {{Catena}}}, \bibinfo {author} {\bibfnamefont {N.}~\bibnamefont
  {{Fornengo}}}, \bibinfo {author} {\bibfnamefont {A.}~\bibnamefont
  {{Masiero}}}, \bibinfo {author} {\bibfnamefont {M.}~\bibnamefont
  {{Pietroni}}}, \ and\ \bibinfo {author} {\bibfnamefont {F.}~\bibnamefont
  {{Rosati}}},\ }\href {\doibase 10.1103/PhysRevD.70.063519} {\bibfield
  {journal} {\bibinfo  {journal} {Phys. Rev. D}\ }\textbf {\bibinfo {volume}
  {70}},\ \bibinfo {pages} {063519} (\bibinfo {year} {2004})},\ \Eprint
  {http://arxiv.org/abs/astro-ph/0403614} {arXiv:astro-ph/0403614 [astro-ph]}
  \BibitemShut {NoStop}%
\bibitem [{\citenamefont {{Zhang}}\ and\ \citenamefont {{Ma}}(2011)}]{zm1}%
  \BibitemOpen
  \bibfield  {author} {\bibinfo {author} {\bibfnamefont {X.}~\bibnamefont
  {{Zhang}}}\ and\ \bibinfo {author} {\bibfnamefont {Y.}~\bibnamefont {{Ma}}},\
  }\href {\doibase 10.1103/PhysRevD.84.104045} {\bibfield  {journal} {\bibinfo
  {journal} {Phys. Rev. D}\ }\textbf {\bibinfo {volume} {84}},\ \bibinfo
  {pages} {104045} (\bibinfo {year} {2011})},\ \Eprint
  {http://arxiv.org/abs/1107.5157} {arXiv:1107.5157 [gr-qc]} \BibitemShut
  {NoStop}%
\bibitem [{\citenamefont {{Zhang}}\ and\ \citenamefont {{Ma}}(2013)}]{zm2}%
  \BibitemOpen
  \bibfield  {author} {\bibinfo {author} {\bibfnamefont {X.}~\bibnamefont
  {{Zhang}}}\ and\ \bibinfo {author} {\bibfnamefont {Y.}~\bibnamefont {{Ma}}},\
  }\href {\doibase 10.1007/s11467-013-0277-0} {\bibfield  {journal} {\bibinfo
  {journal} {Front. Phys. China}\ }\textbf {\bibinfo {volume} {8}},\ \bibinfo
  {pages} {80} (\bibinfo {year} {2013})},\ \Eprint
  {http://arxiv.org/abs/1211.5024} {arXiv:1211.5024 [gr-qc]} \BibitemShut
  {NoStop}%
\bibitem [{\citenamefont {Rovelli}(2004)}]{rov}%
  \BibitemOpen
  \bibfield  {author} {\bibinfo {author} {\bibfnamefont {C.}~\bibnamefont
  {Rovelli}},\ }\href {http://books.google.com.hk/books?id=HrAzTmXdssQC} {\emph
  {\bibinfo {title} {Quantum Gravity}}},\ Cambridge Monographs on Mathematical
  Physics\ (\bibinfo  {publisher} {Cambridge University Press},\ \bibinfo
  {year} {2004})\BibitemShut {NoStop}%
\bibitem [{\citenamefont {Thiemann}(2008)}]{thie}%
  \BibitemOpen
  \bibfield  {author} {\bibinfo {author} {\bibfnamefont {T.}~\bibnamefont
  {Thiemann}},\ }\href {http://books.google.com.hk/books?id=LEBIPgAACAAJ}
  {\emph {\bibinfo {title} {Modern Canonical Quantum General Relativity}}},\
  Cambridge Monographs on Mathematical Physics\ (\bibinfo  {publisher}
  {Cambridge University Press},\ \bibinfo {year} {2008})\BibitemShut {NoStop}%
\bibitem [{\citenamefont {{Ashtekar}}\ and\ \citenamefont
  {{Lewandowski}}(2004)}]{al1}%
  \BibitemOpen
  \bibfield  {author} {\bibinfo {author} {\bibfnamefont {A.}~\bibnamefont
  {{Ashtekar}}}\ and\ \bibinfo {author} {\bibfnamefont {J.}~\bibnamefont
  {{Lewandowski}}},\ }\href@noop {} {\bibfield  {journal} {\bibinfo  {journal}
  {Class. Quant. Grav.}\ }\textbf {\bibinfo {volume} {21}},\ \bibinfo {pages}
  {R53} (\bibinfo {year} {2004})}\BibitemShut {NoStop}%
\bibitem [{\citenamefont {{Han}}\ \emph {et~al.}(2007)\citenamefont {{Han}},
  \citenamefont {{Ma}},\ and\ \citenamefont {{Huang}}}]{hmh}%
  \BibitemOpen
  \bibfield  {author} {\bibinfo {author} {\bibfnamefont {M.}~\bibnamefont
  {{Han}}}, \bibinfo {author} {\bibfnamefont {Y.}~\bibnamefont {{Ma}}}, \ and\
  \bibinfo {author} {\bibfnamefont {W.}~\bibnamefont {{Huang}}},\ }\href@noop
  {} {\bibfield  {journal} {\bibinfo  {journal} {Int. J. Mod. Phys. D}\
  }\textbf {\bibinfo {volume} {16}},\ \bibinfo {pages} {1397} (\bibinfo {year}
  {2007})}\BibitemShut {NoStop}%
\bibitem [{\citenamefont {{Ashtekar}}\ \emph {et~al.}(1998)\citenamefont
  {{Ashtekar}}, \citenamefont {{Baez}}, \citenamefont {{Corichi}},\ and\
  \citenamefont {{Krasnov}}}]{abck}%
  \BibitemOpen
  \bibfield  {author} {\bibinfo {author} {\bibfnamefont {A.}~\bibnamefont
  {{Ashtekar}}}, \bibinfo {author} {\bibfnamefont {J.}~\bibnamefont {{Baez}}},
  \bibinfo {author} {\bibfnamefont {A.}~\bibnamefont {{Corichi}}}, \ and\
  \bibinfo {author} {\bibfnamefont {K.}~\bibnamefont {{Krasnov}}},\ }\href
  {\doibase 10.1103/PhysRevLett.80.904} {\bibfield  {journal} {\bibinfo
  {journal} {Phys. Rev. Lett.}\ }\textbf {\bibinfo {volume} {80}},\ \bibinfo
  {pages} {904} (\bibinfo {year} {1998})},\ \Eprint
  {http://arxiv.org/abs/gr-qc/9710007} {arXiv:gr-qc/9710007 [gr-qc]}
  \BibitemShut {NoStop}%
\bibitem [{\citenamefont {{Ashtekar}}\ and\ \citenamefont
  {{Corichi}}(2003)}]{nonmin2}%
  \BibitemOpen
  \bibfield  {author} {\bibinfo {author} {\bibfnamefont {A.}~\bibnamefont
  {{Ashtekar}}}\ and\ \bibinfo {author} {\bibfnamefont {A.}~\bibnamefont
  {{Corichi}}},\ }\href {\doibase 10.1088/0264-9381/20/20/310} {\bibfield
  {journal} {\bibinfo  {journal} {Class. Quant. Grav.}\ }\textbf {\bibinfo
  {volume} {20}},\ \bibinfo {pages} {4473} (\bibinfo {year} {2003})},\ \Eprint
  {http://arxiv.org/abs/gr-qc/0305082} {arXiv:gr-qc/0305082 [gr-qc]}
  \BibitemShut {NoStop}%
\bibitem [{\citenamefont {{Wang}}\ \emph {et~al.}(2014)\citenamefont {{Wang}},
  \citenamefont {{Ma}},\ and\ \citenamefont {{Zhao}}}]{wmz}%
  \BibitemOpen
  \bibfield  {author} {\bibinfo {author} {\bibfnamefont {J.}~\bibnamefont
  {{Wang}}}, \bibinfo {author} {\bibfnamefont {Y.}~\bibnamefont {{Ma}}}, \ and\
  \bibinfo {author} {\bibfnamefont {X.-A.}\ \bibnamefont {{Zhao}}},\ }\href
  {\doibase 10.1103/PhysRevD.89.084065} {\bibfield  {journal} {\bibinfo
  {journal} {Phys. Rev. D}\ }\textbf {\bibinfo {volume} {89}},\ \bibinfo
  {pages} {084065} (\bibinfo {year} {2014})},\ \Eprint
  {http://arxiv.org/abs/1401.2967} {arXiv:1401.2967 [gr-qc]} \BibitemShut
  {NoStop}%
\bibitem [{\citenamefont {{Wang}}\ and\ \citenamefont
  {{Huang}}(2015{\natexlab{a}})}]{wh2}%
  \BibitemOpen
  \bibfield  {author} {\bibinfo {author} {\bibfnamefont {J.}~\bibnamefont
  {{Wang}}}\ and\ \bibinfo {author} {\bibfnamefont {C.-G.}\ \bibnamefont
  {{Huang}}},\ }\href@noop {} {\  (\bibinfo {year} {2015}{\natexlab{a}})},\
  \Eprint {http://arxiv.org/abs/1505.03647} {arXiv:1505.03647 [gr-qc]}
  \BibitemShut {NoStop}%
\bibitem [{\citenamefont {{Wang}}\ and\ \citenamefont
  {{Huang}}(2015{\natexlab{b}})}]{wh1}%
  \BibitemOpen
  \bibfield  {author} {\bibinfo {author} {\bibfnamefont {J.}~\bibnamefont
  {{Wang}}}\ and\ \bibinfo {author} {\bibfnamefont {C.-G.}\ \bibnamefont
  {{Huang}}},\ }\href {\doibase 10.1088/0264-9381/32/3/035026} {\bibfield
  {journal} {\bibinfo  {journal} {Class. Quant. Grav.}\ }\textbf {\bibinfo
  {volume} {32}},\ \bibinfo {pages} {035026} (\bibinfo {year}
  {2015}{\natexlab{b}})}\BibitemShut {NoStop}%
\bibitem [{\citenamefont {{Huang}}\ and\ \citenamefont {{Wang}}(2015)}]{hw4}%
  \BibitemOpen
  \bibfield  {author} {\bibinfo {author} {\bibfnamefont {C.-G.}\ \bibnamefont
  {{Huang}}}\ and\ \bibinfo {author} {\bibfnamefont {J.}~\bibnamefont
  {{Wang}}},\ }\href@noop {} {\  (\bibinfo {year} {2015})},\ \Eprint
  {http://arxiv.org/abs/1506.02805} {arXiv:1506.02805 [gr-qc]} \BibitemShut
  {NoStop}%
\bibitem [{\citenamefont {{Baez}}(2000)}]{bf1}%
  \BibitemOpen
  \bibfield  {author} {\bibinfo {author} {\bibfnamefont {J.~C.}\ \bibnamefont
  {{Baez}}},\ }\href {\doibase 10.1007/3-540-46552-9_2} {\bibfield  {journal}
  {\bibinfo  {journal} {Lect. Notes Phys.}\ }\textbf {\bibinfo {volume}
  {543}},\ \bibinfo {pages} {25} (\bibinfo {year} {2000})},\ \Eprint
  {http://arxiv.org/abs/gr-qc/9905087} {arXiv:gr-qc/9905087 [gr-qc]}
  \BibitemShut {NoStop}%
\bibitem [{\citenamefont {{Wang}}\ and\ \citenamefont {{Huang}}(2014)}]{wh3}%
  \BibitemOpen
  \bibfield  {author} {\bibinfo {author} {\bibfnamefont {J.}~\bibnamefont
  {{Wang}}}\ and\ \bibinfo {author} {\bibfnamefont {C.-G.}\ \bibnamefont
  {{Huang}}},\ }\href@noop {} {\  (\bibinfo {year} {2014})},\ \Eprint
  {http://arxiv.org/abs/1411.0190} {arXiv:1411.0190 [gr-qc]} \BibitemShut
  {NoStop}%
\end{thebibliography}%
\end{document}